%% file: ictir2023-water-consumption.tex
\documentclass[sigconf,screen]{acmart}
\usepackage{xspace} 
\usepackage{pifont}
\usepackage{hyperref}
\usepackage{enumitem}

\copyrightyear{2023} 
\acmYear{2023} 
\setcopyright{acmlicensed}\acmConference[ICTIR '23]{Proceedings of the 2023 ACM SIGIR International Conference on the Theory of Information Retrieval}{July 23, 2023}{Taipei, Taiwan}
\acmBooktitle{Proceedings of the 2023 ACM SIGIR International Conference on the Theory of Information Retrieval (ICTIR '23), July 23, 2023, Taipei, Taiwan}
\acmPrice{15.00}
\acmDOI{10.1145/3578337.3605121}
\acmISBN{979-8-4007-0073-6/23/07}

\begin{document}
\fancyhead{}
	
\input{sections/abstract}
\input{sections/introduction}

\input{sections/related-work}
\input{sections/analysis}

\input{sections/discussion}
\input{sections/conclusions}

\bibliographystyle{ACM-Reference-Format}
\interlinepenalty=10000
\bibliography{ictir2023-water-consumption}

\end{document}

%% file: sections/abstract.tex
\title{Beyond \COt Emissions: The Overlooked Impact of Water Consumption of Information Retrieval Models}

\begin{abstract}
As in other fields of artificial intelligence, the information retrieval community has grown interested in investigating the power consumption associated with neural models, particularly models of search. 
This interest has become particularly relevant as the energy consumption of information retrieval models has risen with new neural models based on large language models, leading to an associated increase of \COt emissions, albeit relatively low compared to fields such as natural language processing.
Consequently, researchers have started exploring the development of a green agenda for sustainable information retrieval research and operation. 
Previous work, however, has primarily considered energy consumption and associated \COt emissions alone. 
In this paper, we seek to draw the information retrieval community's attention to the overlooked aspect of water consumption related to these powerful models.
We supplement previous energy consumption estimates with corresponding water consumption estimates, considering both off-site water consumption (required for operating and cooling energy production systems such as carbon and nuclear power plants) and on-site consumption (for cooling the data centres where models are trained and operated).
By incorporating water consumption alongside energy consumption and \COt emissions, we offer a more comprehensive understanding of the environmental impact of information retrieval research and operation.
\end{abstract}

\author{Guido Zuccon}
\affiliation{%
	\institution{The University of Queensland}
	\city{Brisbane}
	\country{Australia}}
\email{g.zuccon@uq.edu.au}

\author{Harrisen Scells}
\affiliation{%
	\institution{Leipzig University}
	\city{Leipzig}
	\country{Germany}}
\email{harry.scells@uni-leipzig.de}

\author{Shengyao Zhuang}
\affiliation{%
	\institution{The University of Queensland}
	\city{Brisbane}
	\country{Australia}}
\email{s.zhuang@uq.edu.au}

\ccsdesc[300]{Information systems~Retrieval efficiency}
\ccsdesc[300]{Hardware~Impact on the environment}

\keywords{Green IR, Deep Learning, Water Consumption, IR Sustainability}

\newcommand\todo[1]{{\color{red}#1}}
\newcommand{\COt}{CO$_2$\xspace}
\newcommand{\COtt}{kgCO$_2$e\xspace}
\newcommand{\COte}{kgCO$_2$e/kWh\xspace}
\newcommand{\tick}{\ding{51}}
\newcommand{\cross}{\ding{55}}

\maketitle

%% file: sections/introduction.tex
\section{Introduction}

Over time, information retrieval (IR) systems have increased in complexity, evolving from simple keyword-matching models based on statistics~\cite{robertson2009probabilistic} to feature-derived learning to rank models~\cite{liu2009learning} to the current state of retrieval pipelines that include neural models~\cite{mitra2018introduction,lin2021pretrained,tonellotto2022lecture}. Neural models based on large language models have, in particular, demonstrated exceptional performance in various tasks, such as passage and document retrieval, question-answering, cross-lingual retrieval, and domain-specific search~\cite{xiong2020approximate,lin2021pretrained,zhuang2021fast,formal2021splade,tonellotto2022lecture,zhuang2023augmenting,wang2023can}. However, as models increase in complexity and size, so does their energy consumption~\cite{scells2022reduce}. Strictly associated with energy consumption is the amount of carbon dioxide (\COt) emissions and the \textit{water consumption} entailed by the energy production process and the cooling of the data centers in which these models are executed. These aspects have raised concerns about the environmental impact of information retrieval~\cite{scells2022reduce}.

Previous research has started to address the energy consumption and associated \COt emissions aspects of IR models and research.
In particular, \citet{scells2022reduce} have reported these factors for several popular IR methods and have outlined an agenda for environmentally sustainable IR research. 
However, no attention has been given to the water consumption associated with IR models. Water is a vital resource in this context as it is used for the operation and cooling of energy production systems (i.e., the power plants that provide energy to the data center) and the cooling of data centers in which models are trained and operated.
As water scarcity and drought periods become increasingly pressing global issues~\cite{spinoni2014world,spinoni2018will,zhang2019urban,world2019progress,arthington2018brisbane}, it is essential to evaluate the water consumption of IR models along with their energy consumption and  \COt emissions, to ensure sustainable research and operation.

This paper aims to raise awareness about the water consumption of powerful IR models, providing a comprehensive view of their environmental impact. 
First, we summarize the current state of energy consumption and carbon dioxide emissions in IR research and the green IR agenda put forward in previous work. Next, we discuss the various factors contributing to water consumption in the context of IR models, examining both off-site and on-site consumption aspects. We then move to how water consumption can be measured. Finally, we supplement existing energy consumption estimates of popular IR models with water consumption estimates, and we empirically probe factors that influence water consumption. 

By incorporating water consumption alongside energy consumption and carbon dioxide emissions, this paper aims to promote a broader understanding of the environmental impact of IR research and operation. We encourage researchers and practitioners to consider these factors in their work, ultimately contributing to a more sustainable and responsible approach to developing, researching and deploying IR models.

%% file: sections/related-work.tex
\section{Background and Related Work}
The following gives an overview of the current state of Green AI \& IR and the importance of measuring water in how it contributes to creating meaningful measures of `Green-ness'.

\subsection{Green AI and Green IR}

Concerns about energy consumption in the broader field of artificial intelligence (AI) were heightened after the study from~\citet{strubell2019energy}, which was one of the first to highlight the high energy usage and emissions produced by large language models. This study and others before it~\cite{albers2010energyefficient,belkhir2018assessing} have fueled a growing interest in measuring the energy consumption of research in related fields such as natural language processing (NLP) and machine learning (ML). Within the field of IR, the environmental impacts of technology at scale have been a concern for at least a decade~\cite{chowdhury2012agenda}. 

In general, there are two approaches one can take to quantifying energy and emissions: Life Cycle Assessment (LCA)~\cite{finnveden2009recent} and power consumption measurement. The ISO standard defines LCA as the ``\textit{compilation and evaluation of the inputs, outputs and the potential environmental impacts of a product system throughout its life cycle}''~\cite{iso2006lca}. Due to its complexity and the number of resources required~\cite{chowdhury2012agenda}, most studies choose to measure power consumption directly. Indeed, since then, there have been many efforts to measure energy and emissions for IR systems~\cite{catena2015energy,catena2015study,catena2015loadsensitive,catena2017energyefficient,catena2018efficient,blanco2016exploiting}. Given the explosion in neural models for search based on transformers~\cite{dai2019contextaware,hofstatter2020interpretable,macavaney2020expansion,zhuang2021tilde,qu2021rocketqa,zhuang2021fast,xiong2020approximate,formal2021splade,mallia2021learning,lin2021few}, quantifying the energy and emissions of these systems is more critical than ever. 

Recently,~\citet{scells2022reduce} proposed a framework for minimizing energy usage and emissions for IR with their `reduce, reuse, recycle' methodology. For each concept, they outlined several ways IR practitioners can lower their energy usage: for example, one can \textit{reduce} the number of experiments they do; \textit{reuse} existing pre-trained models for experiments; or \textit{recycle} existing pre-trained models that were initially trained on one task but apply them to another task. Like many other studies before it, this research focused on energy consumption. In this paper, we extend this methodology to consider not only the energy consumption (and, by extension, emissions) but also the water consumption. To the best of our knowledge, our paper marks the first study to investigate the water consumption of IR systems, and neural IR models in particular.

\subsection{Water Consumption in Data Centers}
\label{sec-water-related}
\begin{figure}[t]
	\includegraphics[width=\columnwidth]{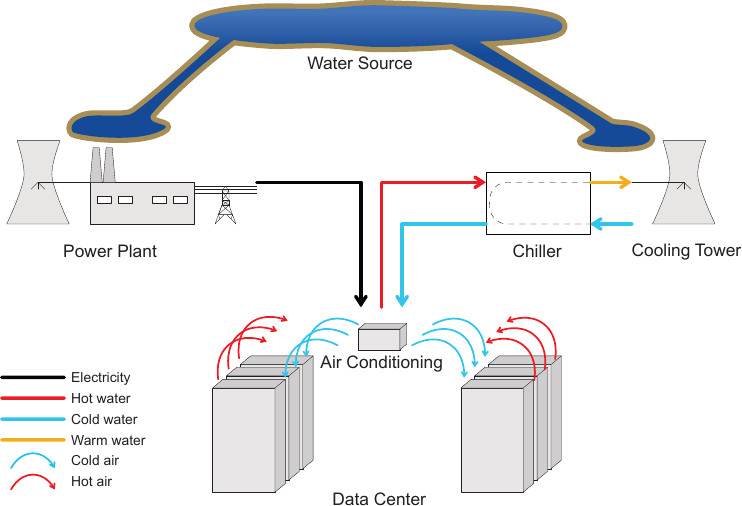}
	\caption{Water usage in a data center and a power plant.}
	\label{fig:water-in-datacenters}
\end{figure}

Water consumption in data centers predominantly occurs due to two distinct factors: first, as an indirect consequence of generating electricity, typically through thermoelectric power sources, and second, as a direct requirement for cooling systems that help maintain the ideal operating environment. These factors are pictured in Figure~\ref{fig:water-in-datacenters}. While water use for electricity generation is well known, its usage for data center cooling is likely less known to IR researchers.

The increase in model complexity associated with the latest advancement of AI models has seen the need for increasingly powerful servers. High-performance servers generate significant amounts of heat, which must be dissipated to maintain optimal operating conditions. Traditional cooling systems often rely on water, using evaporative cooling towers or chilled water systems to maintain temperatures. 
Consequently, the water consumption of these data centers can be significant~\cite{mytton2021data}. 
For example, Google has reported that its data centers consumed 11.4 billion liters of water in the financial year 2017 and 15.8 billion in FY2018\footnote{\url{https://services.google.com/fh/files/misc/google_2019-environmental-report.pdf}.}. The water used to cool data centers is often potable, thus reducing the amount of drinking water available to the population.

The type of cooling system plays a crucial role in determining water usage. For instance, air-cooled systems typically consume less water than water-cooled systems but may be less efficient at dissipating heat. Moreover, the local climate can also impact water consumption; data centers in hot and arid regions may require more water for cooling than those in cooler climates~\cite{karimi2022water}. Finally, seasons and time of day also impact water consumption related to data centers cooling.
The water used in data centers' cooling towers is consumed in two ways: 
\begin{itemize}[leftmargin=*,itemindent=0pt]
	\item through \textit{evaporation}, which occurs as part of the process of cooling, where hot water returning from the data center is sprayed through water distribution nozzles across a cooling fill and then collected at the bottom of the cooling tower in a cold water basis, from where water is pumped back into the chiller connected to the data center's air conditioning; and

	\item through the process of \textit{blow down}, where the water in the pipelines of the data center is flushed. This process of draining water is required to reduce salt, impurity, algae and bacteria concentration, which can cause damage to the cooling system. The higher the water quality, the less blow down of water. 
\end{itemize}

Related to the blow down process is the concept of cycles of concentration $S$: the number of times the dissolved minerals and salts in the circulating water are concentrated compared to the concentration in the makeup water. This concept represents how efficiently the cooling tower system uses and recycles water by measuring how much water is evaporated and concentrated before discharge.
A higher $S$ value indicates that the cooling tower is more efficient in reusing water, reducing overall water consumption and discharge. However, note that as the cycles of concentration increase, so does the concentration of dissolved solids, which can lead to scaling, corrosion, and other issues within the cooling system.

Water consumption related to data centers operations has attracted increasing attention from the research community~\cite{mytton2021data,brocklehurst2021international,karimi2022water}, also in the context of complex and computationally-demanding AI models~\cite{weidinger2022taxonomy,li2023making}. In particular, \citet{li2023making} provide a framework to estimate the water consumption of AI models and investigate the water consumption associated with the training of large language models such as LaMDA~\cite{thoppilan2022lamda}. We build on the framework of \citet{li2023making} to investigate the water consumption related to several common IR models. 

%% file: sections/analysis.tex
\section{Water Consumption Analysis}

In the following sections, we detail a methodology for estimating the water consumption of IR models. We then complement the power and emissions estimates of several well-known IR methods from \citet{scells2022reduce} with the water consumption of these methods. We finally break down these methods' on-site and off-site costs and analyze the influence of factors such as water quality, time of year  and of the day on water consumption.

\subsection{Quantifying Water Consumption in IR}
\label{sec_wcmeasures}
We build upon the framework for measuring the water efficiency of AI models set by~\citet{li2023making}. In this framework, the water consumption $W$ of a model $\mathcal{M}$ is measured as the sum of the water consumed for the cooling of the data center (on-site water consumption, $W_{on}(\mathcal{M})$) and of the water consumed for the production of the electricity used by the data center (off-site water consumption, $W_{off}(\mathcal{M})$):

\begin{equation}
	W(\mathcal{M}) = W_{on}(\mathcal{M}) + W_{off}(\mathcal{M})
	\label{eq_water_consumption}
\end{equation}

The on-site water consumption can then be calculated with respect to the energy $e(\mathcal{M})$ used by the IR model and the water usage effectiveness $WUE_{on}$ of the data center. These can be parameterized by the time energy and water usage occur. This is because water consumption has daytime and season dependencies. Recall that the water is used to cool the data center. In times of the day that are hotter, e.g. in the early afternoons as opposed to the early mornings, or the hotter seasons, e.g., summer as opposed to winter, water consumption will be higher. 

\raggedbottom
We account for this when computing $WUE_{on}$, which then is dependent from $e(t, \mathcal{M})$ and $WUE_{on}(t)$, where $t$ represents a specific time interval (e.g., these quantities could be computed for a 15 minutes interval). Note furthermore that the power consumption of a model may not be constant across all time intervals (e.g., $e(t=i, \mathcal{M}) \ne e(t=j, \mathcal{M})$): this may be the case for example for a model that uses primarily CPU computing in a time interval and GPU computing in another time interval. Given that the water usage effectiveness $WUE_{on}$ of the data center depends on time, running different parts of an IR model pipeline, e.g., different times of the day, may give rise to different water utilization profiles.
Given this, the on-site water consumption is calculated as follows:

\begin{equation}
W_{on}(\mathcal{M})= \sum_{t=1}^{T} e(\mathcal{M}, t) \cdot WUE_{on}(t)  
\end{equation}

\looseness=-1
Water usage effectiveness $WUE_{on}(t)$ typically depends on the cycles of concentration $S$ associated with the blow down of water used in the cooling tower and the outside wet-bulb temperature $T_w$. 

The cycles of concentration for a cooling tower depend on the actual cooling tower specifications and the water quality. As detailed in Section~\ref{sec-water-related}, high-quality water, i.e. with few residues and impurities, requires fewer blow downs of the cooling tower's pipelines. For example, we were able to acquire the cycles of concentration required by two cooling towers of different brands and installed in different locations of the same city in Brisbane, Australia -- cooling tower $A$, located at a private organization, had $S_A=2.25$, while cooling tower $B$, located at a public hospital, had $S_B =1.33$.

\flushbottom
The wet-bulb temperature is the temperature that is measured by a thermometer exposed to the air, and its bulb is covered with a wet wick, which is then exposed to moving air. As water evaporates from the wick, it cools the thermometer bulb, and the temperature reading reflects the cooling effect.
Note that also $T_w$ is dependent on time (of day and season) as it is influenced by factors such as the air temperature, humidity, air pressure, and wind speed. Thus it can be parameterized accordingly, i.e. $T_w(t)$. 
For example, for Brisbane where the previous two cooling towers are located, the mean annual 9am $T_w(t=9am) = 63.5 F$ (min: 53.6F, max: 71.4F), while the mean 3pm $T_w(t=3pm) = 65.3 F$ (min: 57F, max: 72.3F). 

We adopt the same modeling of a cooling tower used by~\citet{li2023making} to compute $WUE_{on}(t)$:

\begin{equation}
	WUE_{on}(t) = \frac{S}{S-1} \cdot \Big (6 \cdot 10^{-5} \cdot T_w(t)^3 - 0.01 \cdot T_w(t)^2 + 0.61 \cdot T_w(t) - 10.40 \Big )
	\label{eq_wue_on}
\end{equation}

Next we consider how to compute the off-site water consumption $W_{off}(\mathcal{M})$.
The off-site water consumption is related to the cooling of the power plant, e.g., in the case of a nuclear or coal power station, and/or the actual production of the electricity, e.g., in the case of a hydroelectric power station. The off-site water consumption could also be null for some electricity production technologies, as it is the case for example for solar power generation. Similarly to the on-site power consumption, also the calculation of $W_{off}(\mathcal{M})$ depends on the energy $e(\mathcal{M})$ used by the IR model, and the water usage effectiveness $WUE_{off}$ -- which in this case refers to the power plant that generates the electricity used by the data center. However, the energy used by the IR model needs to be regulated by the relative amount of energy used by the data center to sustain that power utilization. In other words: for every kWh of electricity used by the components of a server for computation\footnote{These that can be typically measured are CPU, GPU and memory consumption.}, additional energy is used by the data center to power e.g., the storage infrastructure, the power units, the pumps and ventilators used in the cooling system, etc.. This additional consumption is captured by the Power Usage Effectiveness coefficient of the data center, $PUE$. As for the on-site water consumption, all these quantities can be parameterized with respect to time. Thus,  off-site water consumption is calculated as:

\begin{equation}
	W_{off}(\mathcal{M})= \sum_{t=1}^{T} e(\mathcal{M}, t) \cdot PUE(t) \cdot WUE_{off}(t)  
\end{equation}

As mentioned, $WUE_{off}(t)$ is dependent on the power plant(s) used to generate electricity -- and the electricity used may be produced by a mix of fuels (e.g., nuclear, coal, solar). This is modeled as follows. Be $b_k(t)$ the amount of electricity generated using fuel type $k$, and $EWIF_k(t)$ be the Energy Water Intensity Factor, measured in L/kWh, for fuel type $k$. Fro example, typical values for coal are $EWIF_{coal}(t) = 1.7$, and for nuclear $EWIF_{nuclear}(t) = 2.3$~\cite{li2023making}. Then:

\begin{equation}
	WUE_{off}(t) = \frac{\sum_{k} b_k(t) \cdot EWIF_k(t)}{\sum_{k} b_k(t)} 
\end{equation}

Note that typically, and also in the empirical analysis below, values of $k$ and $EWIF_k(t)$ are not known to the researchers for small time intervals $t$: instead, they are more likely able to source an estimate of these values based on yearly reporting from their electricity supplier or government authorities.

\subsection{Water Efficiency of Common IR Models}

Table~\ref{tbl:results} reports the water consumption of common IR models computed according to Equation~\ref{eq_water_consumption} for experiments performed on the MS MARCO-v1 dataset~\cite{nguyen2016msmarco}\footnote{See~\citet{scells2022reduce} for the settings of these experiments. Note, we do not re-run their experiments: we use their values for our water consumption models.}.  
Along with water consumption, we also report running time, power consumption and emissions produced -- these values are sourced from~\citet{scells2022reduce}. To compute water consumption, we used the energy consumption reported by~\citet{scells2022reduce}; note however that in Equation~\ref{eq_water_consumption} $e(\mathcal{M})$ is the Power value in the table divided by the $PUE$ of the data center. 
Furthermore, we used the annual mean wet-bulb temperature at 3pm in Brisbane (65.3 F), cycles of concentration $S=2.25$, a $PUE$ of 1.89\footnote{Which is the $PUE$ of our reference data center, and also that used by~\citet{scells2022reduce}.}, and a combined off-site water usage effectiveness of $WUE_{off}=1.8$, which is representative of that in Brisbane. These results assume time and season in which models are run do not influence water consumption; we consider the impact of these factors in Section~\ref{sec_water_time}. The results show an obvious correspondence between power consumption, emissions, and water consumption: as power consumption increases so do emissions and water consumption. 

\begin{table}[t]
	\small
	\centering
		\caption{Power consumption, emissions and water consumption of IR research over the lifetime of a possible experiment across a number of common IR models. Stages in the pipeline (i.e., model training, indexing, and searching) are reported individually; cumulative values across the full pipeline are shown in bold. Running time, power and emission values are taken from~\citet{scells2022reduce}.}
	\begin{tabular}{p{3.4cm}p{.7cm}p{.7cm}p{.9cm}p{.9cm}}
		\toprule
		Experiment & Running Time (hours) & Power (kWh) & Emissions (\COtt) & Water (L) \\ \hline
		BM25 Indexing & 0.0809 & 0.0021 & 0.0016 & 0.0108 \\ 
		BM25 Search & 0.0025 & 0.00006 & 0.00005 & 0.0003 \\ 
		\cmidrule{2-5} & \textbf{0.0834} & \textbf{0.0022} & \textbf{0.0017} & \textbf{0.0113} \\\midrule
	LambdaMART Training & 0.0285 & 0.0008 & 0.0006 & 0.0041 \\ 
	LambdaMART Rerank + BM25 & 0.0628 & 0.0017 & 0.0013 & 0.0087 \\ 
	\cmidrule{2-5} & \textbf{0.0914} &\textbf{0.0024} & \textbf{0.0019} & \textbf{0.0123} \\\midrule
DPR Training & 16.46 & 6.74 & 5.16 & 34.5910 \\ 
DPR Indexing & 2.42 & 1.04 & 0.7958 & 5.3375 \\ 
DPR Search & 0.4141 & 0.0002 & 0.0001 & 0.0010 \\ 
\cmidrule{2-5} & \textbf{19.3} & \textbf{7.78} & \textbf{5.96} & \textbf{39.9285} \\\midrule
monoBERT Training & 57.43 & 57.95 & 44.38 & 297.4107 \\ 
monoBERT Rerank + BM25 & 23.18 & 10.8 & 8.27 & 55.4277 \\ 
\cmidrule{2-5} & \textbf{80.61} & \textbf{68.75} & \textbf{52.65} & \textbf{352.8384} \\\midrule
TILDEv2 Training & 15.73 & 6.91 & 5.29 & 35.4635 \\ 
TILDEv2 Indexing & 9.44 & 4.74 & 3.63 & 24.3266 \\ 
TILDEv2 Rerank + BM25 & 0.0247 & 0.0007 & 0.0005 & 0.0036 \\ 
TILDE Expansion & 11.89 & 1.04 & 0.7958 & 5.3375 \\ 
\cmidrule{2-5}\vspace{8pt} & \textbf{37.08} & \textbf{12.69} & \textbf{9.72} & \textbf{65.1276} \\\midrule
docTquery Expansion & 760.48 & 169.06 & 129.49 & 867.6489 \\ 
\cmidrule{2-5} & \textbf{785.68} & \textbf{180.71} & \textbf{138.41} & \textbf{927.4389} \\\midrule
uniCOIL Training & 17.97 & 7.24 & 5.54 & 37.1571 \\ 
uniCOIL Indexing & 3.66 & 1.95 & 1.49 & 10.0078 \\ 
uniCOIL Search & 0.8966 & 0.0237 & 0.0182 & 0.1216 \\ 
TILDE Expansion & 11.89 & 1.04 & 0.7958 & 5.3375 \\ 
\cmidrule{2-5}\vspace{8pt} & \textbf{34.41} & \textbf{10.25} & \textbf{7.85} & \textbf{52.6050} \\\midrule
docTquery Expansion & 760.48 & 169.06 & 129.49 & 867.6489 \\ 
\cmidrule{2-5} & \textbf{783.01} & \textbf{178.28} & \textbf{136.54} & \textbf{914.9677} \\\midrule
\end{tabular}
\label{tbl:results}
\end{table}

\begin{figure}[t]
	\includegraphics[width=\columnwidth]{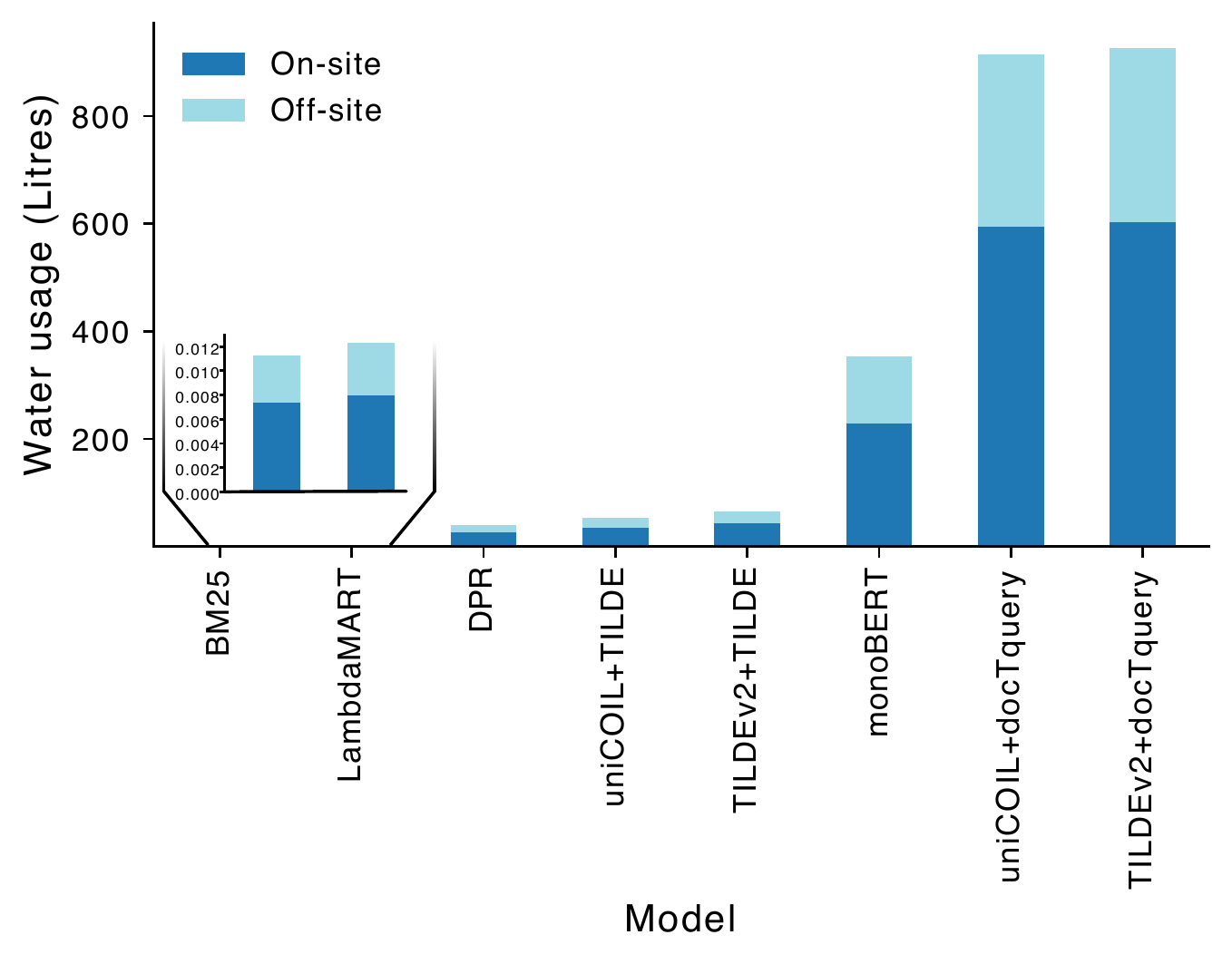}
	\caption{Analysis of on-site and off-site water consumption across different IR models.}
	\label{fig:on-versus-off-site}
\end{figure}

Figure~\ref{fig:on-versus-off-site} breaks down the water consumption of the considered IR models with respect to on-site and off-site consumption. This figure shows that for each model, under the considering values of the parameters, water consumption is dominated by the need of cooling the data center. This result is highly influenced by (1) the specifications and quality of the cooling tower, which influence the second part of Equation~\ref{eq_wue_on}, (2) the quality of the water used for cooling, which in turns influences the cycles of concentrations $S$, (3) and the wet-bulb temperature measured at the location of the data center, which in turns is influenced by both the local climate and the season and time of day in which computations occur. 



\label{sec_water_time}

\begin{figure}[t]
	\includegraphics[width=\columnwidth]{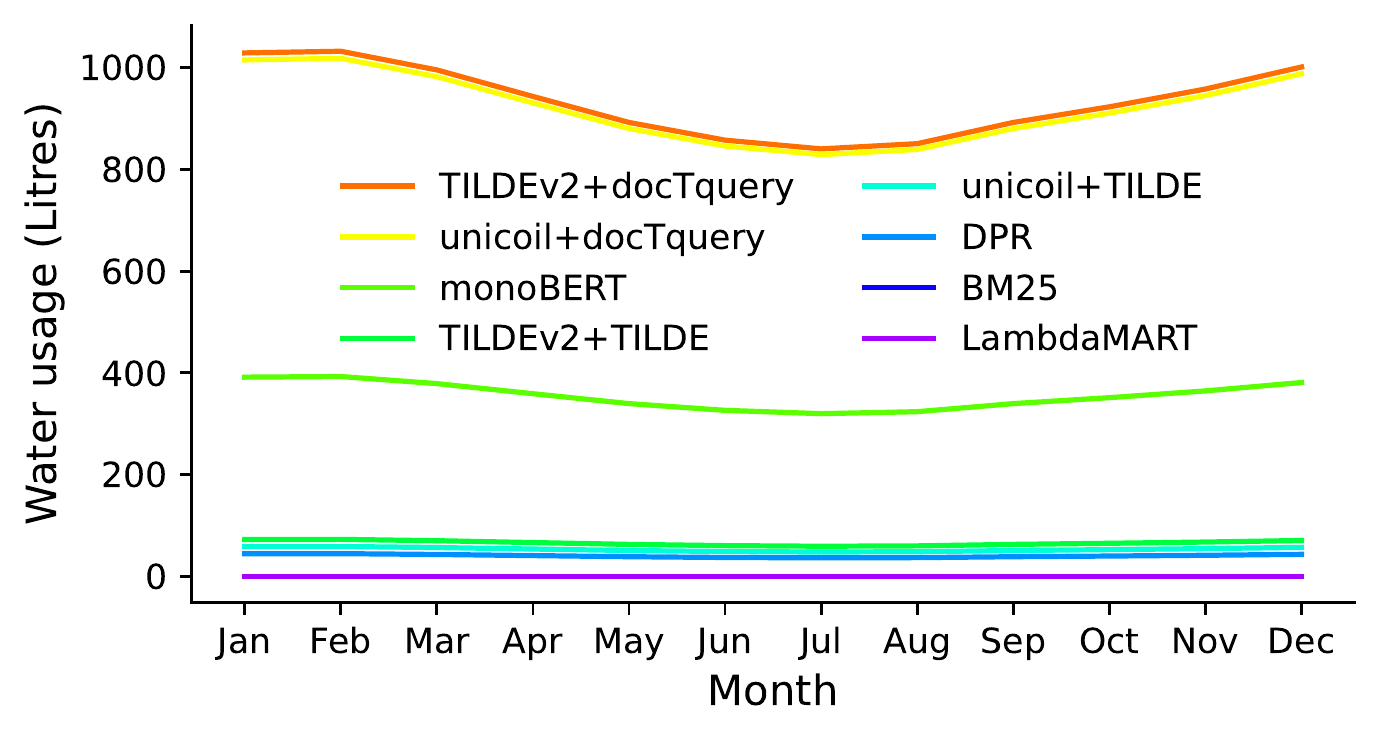}
	\caption{Water consumption of the considered IR models throughout a year for Brisbane, Australia (thus November-February is summer). }
	\label{fig:water-efficiency-year}
\end{figure}

\subsection{Effect of Water Quality on Water Usage}
We demonstrate the influence of cycles of concentration, and thus water quality, on on-site water consumption through an example. Recall that the worse the water quality, the more the sediments in the water and thus the need to blow down (i.e. ``flush'') the water pipelines of the cooling system of the data center to avoid damages.

\raggedbottom
For this example we have considered the TILDEv2 model~\cite{zhuang2021fast} with docTquery expansion~\cite{nogueira2019doc2query}; other models show similar trends, although this was the IR model with the largest water consumption in the results of Table~\ref{tbl:results}.
We have computed the on-site water consumption ($W_{on}$) using the values of cycle of concentrations from the two cooling towers A and B mentioned in Section~\ref{sec_wcmeasures}, where A was located in a private organization and B in a public hospital ($S_A=2.25$, $S_B=1.33$): cooling tower $A$ is more water efficient than $B$. We keep all other values the same as those used for Table~\ref{tbl:results}. For site $A$, we obtain $W_{on,A}(TILDEv2) = 602.1609$ L, while for site $B$, we obtain $W_{on,B}(TILDEv2)=1261.3524$ L. This example materializes the large impact a lower water quality can have on the amount of water consumed by a data center. The use of high quality water supplies however constitutes an issue per se. First, chemicals could be used to improve water quality: but these add extra costs in the running of the data center operation. These chemicals may also have secondary negative effects on the environment if released into natural water-streams at blow down. Chemicals typically used in cooling towers in fact include corrosion and scale inhibitors, algaecides and biocides, and pH adjusters, which may have harmful impact on the environment~\cite{goni2019green} -- though we note environmental friendly products do exist, e.g., green corrosion inhibitors~\cite{goni2019green}. 
 Second, potable water, i.e. drinking water, is typically of high quality -- using this water for cooling means subtracting drinking water to the local population, which is problematic in areas with scarce access to sources of potable water.

\begin{figure}[t]
	\includegraphics[width=\columnwidth]{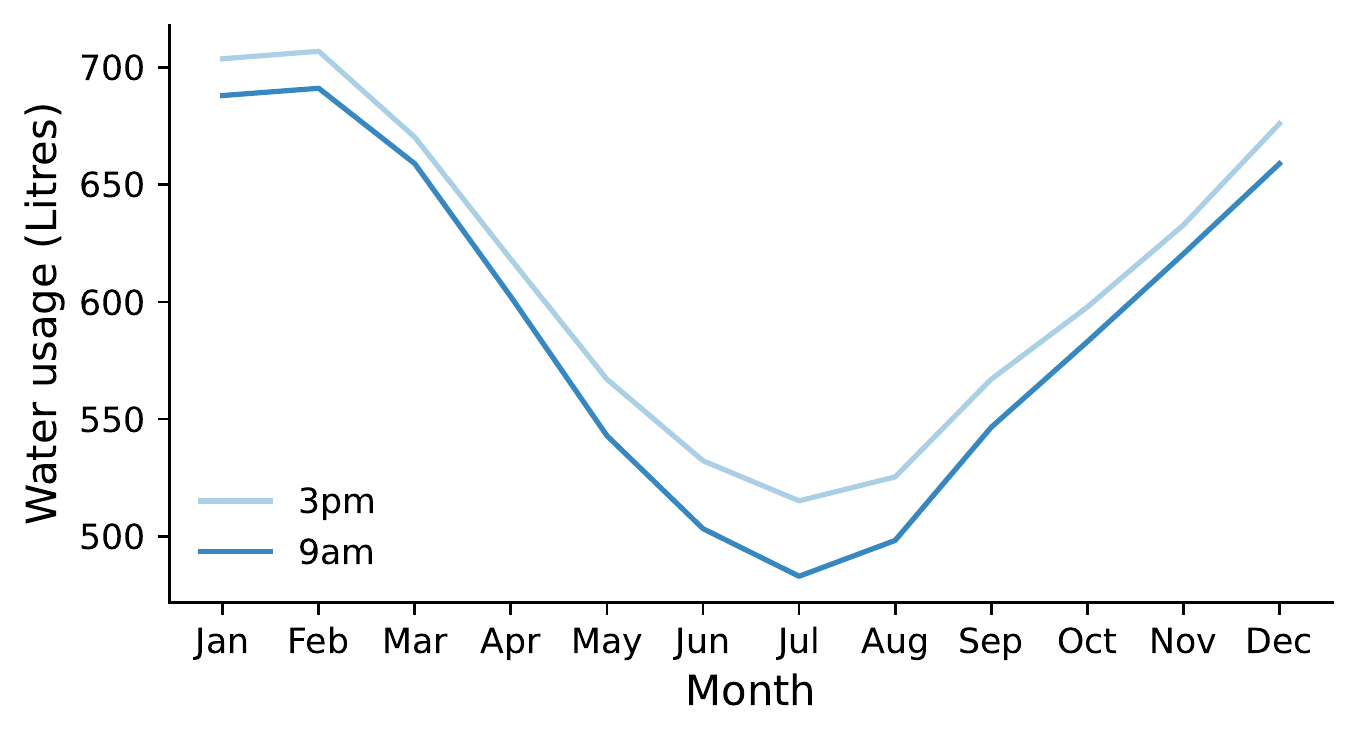}
	\caption{Comparison between on-site water consumption $W_{on}$ for TILDEv2 with docTquery expansion obtained when running the model in the morning (9am) vs. in the afternoon (3pm). The analysis is performed for each month of the year.}
	\label{fig:water-efficiency-day}
\end{figure}

\subsection{Effect of Time on Water Usage}
\label{sec-time-day}
Next we discussion the effect of time of day and season on water consumption. This effect is due to the rise of the wet-bulb temperature that is typically associated to the warmer part of the day, e.g., afternoons, or to the warmer seasons, e.g., summer. This can be observed clearly in Equation~\ref{eq_wue_on}: as the temperature $T_w(t)$ raises, so does $WUE_{on}(t)$ and thus consequently $W_{on}$. 

To materialize the effect of temperature changes on (on-site) water consumption, in Figure~\ref{fig:water-efficiency-year} we report the total water usage $W$ of each IR model throughout the year. We base these results on the same settings used to generate the estimations in Table~\ref{tbl:results}, but instead of taking the annual mean wet-bulb 3pm temperature for Brisbane, we consider the \textit{monthly} mean wet-bulb 3pm temperature. Note that Brisbane is in the Southern Hemisphere, and thus summer is in the period December through to February, and winter is in the period June through August. The figure shows that for the most ``thirsty'' IR model considered, TILDEv2 with docTquery expansion, water consumption can vary of  192 liters: running this model in the hottest month consumes $23\%$ more water than running it in the coolest month. On the other hand, the impact of season on water consumption for models like BM25 and LambdaMART is minimal (at least in absolute terms). 

We perform a similar analysis for showing the impact of time of day. For this, we limit the analysis to the model with highest water consumption, TILDEv2 with docTquery expansion, as an example. We consider two times of the day, 9am and 3pm, and the month with largest difference in mean temperatures at these times in Brisbane: July. The mean 9am wet-bulb temperature in July is 53.6F, and at 3pm is 57.02F: thus the difference is 3.42F. Assuming all other variables are set to the values used for Table~\ref{tbl:results}, the on-site water consumption of TILDEv2 with docTquery expansion at 9am is $W_{on}(TILDEv2,July@9am) = 482.90$, while at 3pm is $W_{on}(TILDEv2,July@3pm) = 515.1$: a difference of 32.2 liters. Figure~\ref{fig:water-efficiency-day} extends the comparison between the mean 9am and 3pm on-site water consumption for TILDEv2 with docTquery to all months (for each month, we consider the mean over the last 16 years\footnote{Due to the availability of this data from the local meteorology agency.}). The figure suggests that for Brisbane the difference in water consumption obtained when running the model at 9am instead than at 3pm is higher during the winter months, than during the summer months. This is because in Brisbane temperature differences between mornings and afternoons are higher in winter than in summer. We note that these findings are location-dependent.

%% file: sections/discussion.tex
%
%

%% file: sections/conclusions.tex
\section{Discussion and Conclusion}

AI and IR models naturally consume electricity and, as a result, may produce emissions. However, electricity alone is not the only limited resource these models consume. In training or operating IR models, water usage is an important factor to take into consideration, especially when based on increasingly computationally demanding large neural models. Globally, water is increasingly becoming a scarce resource, especially high-quality, potable water\footnote{\url{https://www.un.org/en/climatechange/science/climate-issues/water}.}. 

A reduction in energy or emissions does not necessarily translate into a reduction in water consumption. This is because, in the context of IR models, water consumption does not only depend on the amount of energy consumed (and thus heat generated that needs to be dissipated through cooling) but also on the temperature, humidity and wind conditions of the environment, which naturally fluctuate with times of day and seasons. This aspect is important because it suggests that strategies that reduce \COt emissions do not necessarily obtain a reduction of water consumption: indeed, the opposite may occur. This is the case for example for solar power: while this method of electricity production emits no \COt, it is most efficient in times of the day and seasons with high solar irrigation -- which in turn are associated with high temperatures and thus higher quantity of water required to cool the data centers (Section~\ref{sec-time-day}).

With this paper, we aim to raise awareness about the water consumption associated with large IR models so that practitioners can be conscious of their impact and take steps to minimize water consumption. To this end, we have presented a method for quantifying the water usage of IR methods and compared several models in terms of not only their energy usage and emissions production, but also in terms of their water usage. We have shown that while water consumption of keyword-matching and learning to rank models is minimal, the water consumption associated with neural models can be significant. 
The analysis we have provided in this paper comes with limitations, first and foremost because power, water and emissions calculations were performed using estimated energy usage and average temperatures. 

\looseness=-1
To further help the IR community to be conscious of the impact of their research on the environment and monitor their energy and water consumption along with their \COt emissions, we have developed: (1) a web-based calculator where researchers can conveniently insert the parameters associated to their models and data centers, and compute \COt emissions and water consumption, and (2) a plug-in for the Weights \& Biases telemetry tool\footnote{\url{https://wandb.ai/}, an MLOps tool for performance visualization and experimental tracking of machine learning models.} that allows tracking energy and water consumption, and \COt as experiments are run. This material along with other resources associated to the paper are available at \url{https://github.com/ielab/green-ir}.

\subsubsection*{Acknowledgments}
This research is funded by the Australian Research Council Discovery Project DP210104043.